\documentclass[fleqn,twoside]{article}
\usepackage{espcrc2}
\usepackage{graphicx}
\usepackage[figuresright]{rotating}
\def\fun#1#2{\lower3.6pt\vbox{\baselineskip0pt\lineskip.9pt
\ialign{$\mathsurround=0pt#1\hfil##\hfil$\crcr#2\crcr\sim\crcr}}}

\title{Photons, Clocks, Gravity and the Concept of Mass}
\author{L.B.Okun\address{ITEP, Moscow, 117218, Russia}
 \thanks{email: okun@heron.itep.ru}}
\date{}

\begin{document}

\begin{abstract}

Transparencies of the 18th Henry Primakoff lecture at the University of
Pennsylvania, Philadelphia, USA, April 11, 2001 and of the Special lecture
given at the 7th International Workshop on Topics in Astroparticle and
Underground Physics, TAUP 2001, Laboratori Nazionali
del Gran Sasso, Italy, September 9, 2001; to be published by
Elsevier Nucl. Phys. B (Proceedings supplements).

Various aspects of the concept of mass in modern physics are
discussed: 1. Mass in Special Relativity, 2. Atoms in Static
Gravity, 3. Photons in Static Gravity, 4. Misleading Terminology,
5. Unsolved Problems of mass.

\end{abstract}

\maketitle

\newpage

\section{MASS IN SPECIAL RELATIVITY}
\begin{enumerate}
\item The term mass was introduced by Newton in {\it Principia},
1687: ``Definition I: The quantity of matter is the measure of the
same, arising from its density and bulk conjointly''.
\item The principle of relativity was formulated by Galileo in
{\it Dialogue}, 1632 (Galileo's ship).
\item His predecessor: Nicolaus Cusanus in ``De docta ignorata''
(``On the scientific ignorance''), 1440.
\item Velocity of light $c$ was first measured by O.Roemer, 1676.
\item A. Michelson and E. Morley, 1887: \\
$c$ is the maximal velocity of signals. $c = 1$.
\item Special relativity: Interferometer of Michelson on the ship
of Galileo.
\item Lorentz transformations of $t$, ${\rm \bf r}$ and $E$,
${\rm\bf p}$, where  ${\rm \bf r}$ and ${\rm \bf p}$  are
3-vectors.
\item $m^2 = E^2 -{\rm\bf p}^2$ ; $E$, ${\rm\bf p}$ -- 4-vector;
$m$ -- scalar.
\item Einstein 1905: rest energy $E_0 =m$.
\item
$ E = m\gamma \; , \;\; {\rm\bf p} = m\gamma {\rm\bf v} = E{\rm\bf
v} \;, \;\; {\rm\bf v} =\frac{\rm\bf p}{E} $ $$ v=|{\rm\bf v}| \;
, \;\; \gamma = 1/\sqrt{1-v^2} \; , \;\; v\leq 1 \;\; .
$$
\item Photon: $m=0$, $E=p$, $v=1$, no rest frame.
\item System of two bodies:
$$
E=E_1 +E_2 \; , \;\; {\rm\bf p} = {\rm\bf p}_1 + {\rm\bf p}_2  \;\; ,
$$
$$
m^2 = E^2 - {\rm\bf p}^2 = (E_1 +E_2)^2 -({\rm\bf p}_1 +{\rm\bf p}_2)^2 =
$$
$$
= m_1^2 + m_2^2 + 2E_1 E_2 (1-{\rm\bf v}_1{\rm\bf v}_2) \;\; .
$$

Two photons: $m_{2\gamma}^2 = 2E_1 E_2(1-\cos\theta)$.
\item Mass of a system is conserved. \\
Example $e^+ e^- \to 2\gamma$ at rest. \\
$m_{e^+ e^-} = m_{2\gamma} = 2m_e$, though $m_{\gamma} =0$.
\item Mass is not a measure of inertia. \\
This measure is derived from: $d{\rm\bf p}/dt = {\rm\bf F}$, ${\rm\bf p} =
E{\rm\bf v}$.
\item Mass is not additive at $v\neq0$. \\
$m=m_1 +m_2$ only at $v_1 = v_2 =0$.
\item Mass is not a source and receptor of gravity  at $v\neq 0$
(see item 20).
\item Mass is not a measure of an amount of matter. The concept of
matter is much broader in relativity
than in classical physics (examples: gas or flux of photons
and/or neutrinos).
\item From mathematical point of view mass occupies a higher rank
position in special relativity than in non-relativistic physics:
it is a relativistic invariant.
\item From practical point of view the role of mass of a particle
is decreasing with increase of its energy.

\section{ATOMS IN STATIC GRAVITY}

\item According to General Relativity, the source and receptor
of gravitational field  is the density of energy-momentum
tensor $T^{ik}$ which is coupled to gravitational field like
density of electric current is coupled to electromagnetic field.

For a point-like particle $$ T^{ik} = m\delta({\rm\bf r})
\frac{dx^i}{d\tau} \frac{dx^k}{dt} \;\; , \;\;
\tau = \sqrt{t^2 -{\rm\bf r}^2} \;\; .
$$
For a particle at rest ($v=0$) mass $m$ is the source and
receptor of gravity: $T^{00} = m\delta({\rm\bf r})$.

Thus, according to GR, there is no such notion as gravitational
mass $m_g$.
\item $\phi$ -- gravitational potential
\begin{description}
\item{a)} the potential of sun:
$$ \phi = -\frac{GM_{\odot}}{rc^2} = -\frac{1}{2} \frac{r_g}{r} \;
,  \;\; \phi(\infty) = 0
$$
$G$ -- Newton constant,
 $M_{\odot}$ -- mass of the sun, \\
  gravitational radius $r_g = 2GM_{\odot}/c^2 = 3$ km.
\item{b)}
the potential near the surface of earth at height $h$:
$$
\phi = \frac{gh}{c^2} \; , \;\; \phi(0) = 0 \;\; , g = 9.8 \; {\rm m~ s}^{-2} \;\; .
$$
\end{description}
\item $E_{0g} = m + m\phi = m(1+\phi)$, $E_{0g} \neq m$. \\
$E_{0g}$ -- rest energy in a static gravitational field.  It
increases with increasing distance from the source.
\item For an excited level with mass $m^*$
$$ E_{0g}^* = m^* (1+\phi) \;\; . $$
The splitting between levels
increases with $h$:
$$ \Delta E_{0g} = E_{0g}^* - E_{0g} = (m^* -m)(1+\phi) \;\; . $$
\item Note that for electric potential the splitting is constant:
$$
E_{0e} = m^* +e\phi_e \;\; .
$$
Consider an excited ion He$^+$ in a capasitor:
$$
\Delta E_{0e} = E_{0e}^* - E_{0e} = m^* -m \;\; .
$$
\item The frequency $\omega$ of a photon emitted in the process
of deexcitation
$$
\omega = \Delta E_{0g}/h \;\; ,
$$
where $h=2\pi\hbar$ is the Planck constant. \\
Hence atomic clocks in gravitational field
increase their rate with height, as
predicted by Einstein in 1907.
\item This  was confirmed by airplane experiments in 1970s. \\
Feynman: ``atoms at the surface of the earth are a couple of days
older than at its center''.
\item Thought experiment: Carry clock A from the first floor to
the second.  In a year carry clock B the same way.
 A will be ahead of B. $\Delta T/T = gh/c^2$.
Absolute effect! \\
This thought experiment is used to get rid of disparity between
a clock on an airplane and that on a desk. (Think of twin
paradox in special relativity.)
\item Pendulum clocks are not standard clocks similar to
wristwatches, or atomic clocks.\\
 $T\sim \sqrt{l/g}$. \\
 They are gravimeters,
measuring the strength of gravitational field.

\section{PHOTONS IN STATIC GRAVITY}

\item In 1911 Einstein predicted redshift of photon frequency in
static gravitational field: the gravitational redshift.
\item In 1960 Pound and Rebka discovered the shift $\Delta \omega
/\omega = gh/c^2$ in $~^{57}{\rm Fe}$. Interpretation in terms of
``the weight of the photon''.
\item For static field and static observers the frequency of photon does not
depend on height. (Maxwell equations in static field).
\\ Hence the shift observed by Pound et al. was due to the larger
splitting of $~^{57}{\rm Fe}$ levels at the top of the Harvard
tower. Redshift of photons relative to clocks.
\item Unlike frequency, the momentum of photon decreases with height. \\
Schwarzshild metric in general relativity $g^{ij}$. \\
 The masslessness of photon:
$$ g^{ij}p_i p_j =0  , \; i,j = 0,1,2,3  , \; g^{00}p_0^2
-g^{rr}p_r^2 = 0 $$ $$ g^{00} =1/(1+ 2\phi) \; , \;\; g^{rr} =
(1+2\phi) \;\; , $$
$p_0 \equiv E$, hence, $$ p_r = \frac{E}{1+2\phi} =
\frac{E}{1- r_g/r} \;\; . $$
\item Photon's wave length $\lambda$ increases with height. In
that sense photon is redshifted $$ \lambda = \frac{2\pi
\hbar}{p_r} = \frac{2\pi \hbar}{E}(1+2\phi) $$
\item Coordinate velocity is smaller near the sun:
$$
v=\frac{\lambda \omega}{2\pi} =(1+2\phi)
$$
\item Another derivation: Consider interval
$$
ds^2 = g_{00}dt^2 -g_{rr}dr^2 =0
$$
$$
g_{00} =\frac{1}{g^{00}} = 1+ 2\phi \; , \;\; g_{rr} = \frac{1}{g^{rr}} = \frac{1}{1+2\phi}
$$
$$
v=\frac{dr}{dt} = \sqrt{g_{00}/g_{rr}} = 1+ 2\phi = 1-\frac{r_g}{r} < 1
$$
\item Before general relativity (1906, 1911) Einstein had $v= 1+\phi$
(there was no $g_{rr}$ at that time).
In 1915: $(1+ 2\phi)$.
\item Geometrical optics. Refraction index
$$ n=\frac{1}{v} =\frac{1}{1+2\phi} \approx 1+ \frac{r_g}{r} \; ;
\;\; n-1 = \frac{r_g}{r} \;\; . $$
\item Deflection of star light by sun (1911, 1915, 1919)
$$ \alpha = \frac{2r_g}{R_{\odot}} \sim 10^{-5} $$
\item Gravitational lensing of galaxies.
\item Delay of radar echo from a planet in upper conjunction
(Shapiro, 1964)
$$
\Delta t = \frac{2}{c} \int\limits^{z_p}_{z_e} dz
(\frac{1}{v}-1) = \frac{2}{c} \int\limits^{z_p}_{z_e} dz
\frac{r_g}{r} =
$$
$$
= 2\frac{r_g}{c} \ln \frac{4r_p
r_e}{R_{\odot}^2} \approx 240 \mu {\rm s}
$$
Earth: $r_e = 150 \cdot
10^6$ km. \\
Mercury: $r_p = 58 \cdot 10^6$ km, \\
 $R_{\odot} = 0.7 \cdot 10^6$ km.
\item  Deceleration  of a particle approaching the sun is a
relativistic effect, characteristic not only to photons, but to
any fast enough particle. M.I. Vysotsky (private communication)
has shown that ``enough'' means $v_\infty > c/\sqrt{3}$, where
$v_\infty$ is velocity of the particle at $r=\infty$.

To prove this consider the expression for energy of a particle of
mass $m$ falling on the sun along radius $r$ with coordinate
velocity $v = dr/dt$ (use eq. (88.9) from ``Field Theory'' by L.
Landau and L. Lifshitz): $$ E =
\frac{mc^2\sqrt{g_{00}}}{\sqrt{1-\frac{g_{rr}}{g_{00}}v^2}} \;\; .
$$ Note that, according to Schwarzschild, $g_{00} =
\frac{1}{g^{rr}} = 1- r_g/r$ and that energy does not depend on
$r$. Hence $$ 1-v_\infty^2 = g_{00}^{-1} - g_{00}^{-3}v^2 \;\; .
$$ For $r_g/r \ll 1$ one gets $$ v^2 = v_\infty^2 +
r_g/r(1-3v_\infty^2) \;\; . $$ Thus, the particle accelerates like
a non-relativistic stone only if $v_\infty < c/\sqrt{3}$.

\section{MISLEADING TERMINOLOGY}

\item If, instead of ${\rm \bf p}= E{\rm\bf v}/c^2$, the non-relativistic
formula ${\rm\bf p}=m{\rm\bf v}$ is kept, then $m=E/c^2$.
This ``relativistic mass'', often denoted as $m_r$,
is another notation for energy.
\item When applied to photon ``$m_\gamma = E_\gamma/c^2$''. \\
Poincare, Lorentz, Born, Pauli, partly Einstein.
\item  Einstein coined the expression ``energy - mass
equivalence'',
which sometimes he used in the sense ``whenever there is energy,
there is mass''. (Recall massless photon.)
\item A redundant term ``rest mass'' $m_0$ is abundantly used.
\item Gravitational redshift is interpreted as ``loss of photon's energy as it
climbs out of gravitational potential''.
\item From the point of view of relativity no $m_r$, $m_0$, $m_g$,
$m_i$, $m_l$, $m_t$. Only $m$!

\section{UNSOLVED PROBLEMS OF $\mbox{\boldmath$m$}$}

\item Is the Higgs mechanism correct? \\
If yes, we have to discover higgses and to study the pattern of
their couplings.
\item Can the hierarchy problem ($m_{Pl} \simeq 10^{19}$ GeV {\it vs}
$m_Z \simeq 100$ GeV) be solved by SUSY? \\ If yes, we have to
discover numerous superparticles and to ascertain the pattern of
their masses and couplings.
\item SUSY, if confirmed, would lead us into the quantum
dimensions of space and time, into superspace.
\item Supergravity might lead to superunification of all forces.
\item The scale and mechanism of SUSY breaking is of paramount
importance.
\item What makes the critical energy density of vacuum
$\varepsilon_c \simeq 10^{-47}$ GeV$^4$ so small compared with
$\eta^4$, the density of vacuum expectation value (VEV) of the Higgs
field: $\eta = (246$ GeV)?
\item If the recent data on the universe expansion are correct,
how to explain the pattern of cosmological energy densities of
baryons of dark matter, and of vacuum: $$ \Omega_B \equiv
\varepsilon_B/\varepsilon_c \simeq 0.03  , \;
 \Omega_{DM} \equiv
\varepsilon_{DM}/\varepsilon_c \simeq 0.3 \; , $$ $$
\Omega_{\Lambda} \equiv \varepsilon_{\Lambda}/\varepsilon_c \simeq
0.7 \;\; ? $$
\item Compared with the above enigmas, the problem of neutrino
masses and mixings may seem to be not so lofty. But it could
reveal important clues to the concept of mass.
\item A serious twist to the concept of mass was given by quarks
and gluons. The basic definition $m^2 = E^2 - {\rm \bf p}^2$ is not
applicable to them, as they never exist as free particles with
definite value of momentum. This colored particles are confined
within colorless hadrons.
\item That reminds us that when speaking about mass of a particle
we have to specify the distance (momentum tranfer $q$) at which it
is measured. Masses as well as charges are running functions of
$q^2$. It is not enough to write down the fundamental Lagrangian.
One has to indicate the scale to which its couplings and masses
refer.
\item At short distances the masses of light $u$-quarks and
$d$-quarks, the building blocks of nucleons, are smaller than 10
MeV, while masses of nucleons are hundred times larger. These
short-distance quarks are called ``current quarks''.
\item The same quarks with their ``gluon coats'' are called
``constituent quarks'', their masses being about 300 MeV. There
should exist ``constituent gluons''. Hadrons without ``valent''
quarks are called ``glueballs'', hadrons containing both ``valent''
quarks and gluons are referred to as ``hybrids''.
\item The quark-antiquark and gluonic vacuum condensates: $<q \bar
q>$, $<GG>$, etc. play an important role.
The masses of hadrons appear due to ``burning
out'' of these (negative energy density) vacuum condensates inside
hadrons by the intense color fields of the valent constituents.
Though this picture has not been rigorously proved, it seems
evident that the origin of mass is in vacuum.
\item The existing terminology is obsolete and non-adequate.
Fermions (quarks and leptons) are usually referred to as
``matter'', though neutrinos kinematically behave like photons.
Even very heavy bosons are often referred to as ``radiation''.
Actually all fields and their quantum excitations (including
photons) are matter.

Massive vacuum; is it also matter?
\item The literature on the subjects discussed in this lecture is
very vast and controversial. The interested reader might look for
further references in article \cite{1,2,3,4,5,6,7,8,9,10,11,12}
\end{enumerate}

\section*{ACKNOWLEDGEMENTS}

I am grateful to Paul Langacker, Alfred Mann and their colleagues
at the Physics Department of University of Pennsylvania and also
to Alessandro Bettini and Venya Berezinsky at LNGS for the kind
invitation to give this lecture and  for their warm hospitality.
The work was partially supported by A. von Humbold Award and by
RFBR grant No. 00-15-96562.


\begin{thebibliography}{99}
\bibitem{1}
 On the photon mass (with I.Yu.Kobzarev). UFN {\bf 95}
 (1968) 131-137; \\ Sov. Phys. Uspekhi {\bf 11} (1968) 338-344.
\bibitem{2}
 The concept of mass.  Physics Today, June 1989, pp.
31-36
\bibitem{3}
Putting to rest mass misconseptions.  Physics Today, May 1990,
pp.13,15,115-117.
\bibitem{4}
The Problem of Mass: From Galilei to Higgs.   Opening lecture at
the 29th Intern. School of Subnuclear Physics, in "Physics at the
Highest Energy and Luminosity to Understand the Origin of Mass",
v.29, Ed. A.Zichichi (Plenum Press, N.Y., 1992) pp.1-24.
\bibitem{5}
Note on the meaning and terminology of Special Relativity.   Eur.
J. Phys. {\bf 15} (1998) 403-406.
\bibitem{6}
Comment on the physical meaning of '$c$' in relativistic equations
 (with B. Webber).
Eur. J. Phys. {\bf 20} (1999) L47.
\bibitem{7}
 Gravitation, Photons, Clocks
(with K. Selivanov, V.Telegdi). UFN  {\bf 169} (1999) 1141-1147;
Phys. Uspekhi {\bf 42} (1999) 1045-1051.
\bibitem{8}
On the interpretation of the redshift in a static gravitational
field (with K. Selivanov, V. Telegdi). Am. J. Phys. {\bf 68}
 (2000) 115-119.
\bibitem{9}
Photons and static gravity.   Mod. Phys. Lett. {\bf A15}
 (2000) 1941-1947; preprint hep-ph/0010120.
\bibitem{10}
A thought experiment with clocks in static gravity. Mod. Phys.
Lett. {\bf A15} (2000) 2007-2009; preprint hep-ph/0010256.
\bibitem{11}
Reply to the letter ``What is mass?'' by R.I.Khrapko. UFN {\bf
170} (2000) 1366-1371; Phys. Uspekhi {\bf 43} (2000) 1270-1275.
\bibitem{12}
Relation between energy and mass  in  Bohr's essay on his debate
with Einstein.
 Yad. Fiz. {\bf 64} (2001) 590-593;
 Phys. Atomic Nuclei {\bf 64} (2001) 536-539.
\end{thebibliography}
\end{document}